\documentclass[a4paper,11pt]{amsart}
\usepackage{graphicx}
\usepackage{amssymb}
\begin{document}

\hyphenation{gra-vi-ta-tio-nal re-la-ti-vi-ty Gaus-sian
re-fe-ren-ce re-la-ti-ve gra-vi-ta-tion Schwarz-schild
ac-cor-dingly gra-vi-ta-tio-nal-ly re-la-ti-vi-stic pro-du-cing
de-ri-va-ti-ve ge-ne-ral ex-pli-citly des-cri-bed ma-the-ma-ti-cal
de-si-gnan-do-si coe-ren-za pro-blem gra-vi-ta-ting geo-de-sic
per-ga-mon cos-mo-lo-gi-cal gra-vity cor-res-pon-ding
de-fi-ni-tion phy-si-ka-li-schen ma-the-ma-ti-sches ge-ra-de
Sze-keres con-si-de-red tra-vel-ling ma-ni-fold re-fe-ren-ces
geo-me-tri-cal in-su-pe-rable sup-po-sedly at-tri-bu-table
Bild-raum in-fi-ni-tely counter-ba-lan-ces iso-tro-pi-cally
pseudo-Rieman-nian cha-rac-te-ristic geo-de-sics
Koordinaten-sy-stems ne-ces-sary ge-ne-ra-ted sym-me-tri-cal}

\title[Proof of a Lorentz and Levi-Civita thesis] {{\bf Proof of a Lorentz and Levi-Civita thesis}}

\author[Angelo Loinger]{Angelo Loinger}
\address{A.L. -- Dipartimento di Fisica, Universit\`a di Milano, Via
Celoria, 16 - 20133 Milano (Italy)}
\email{angelo.loinger@mi.infn.it}

\vskip0.50cm

\begin{abstract}
A formal proof of the thesis by Lorentz and Levi-Civita that the
left-hand side of Einstein field equations represents the real
energy-momentum-stress tensor of the gravitational field.
\end{abstract}

\maketitle

\vskip1.20cm \noindent \small \textbf{Summary} -- \textbf{1}.
Introduction. Aim of the paper. -- \textbf{2}. Mathematical
preliminaries. -- \textbf{3}. Proof that the left-hand side of the
Einstein field equations gives the true energy-momentum-stress
tensor of the gravitational field. -- \textbf{4}. A fundamental
consequence. -- \emph{Appendix}: On the pseudo energy-tensor.

\normalsize

\vskip1.20cm \noindent \textbf{1.} -- As it has been remarked
\cite{1}, if $I$ is the \emph{action} integral of any field (of
any tensorial nature) -- say $\varphi(x)$, $x \equiv (x^{0},
x^{1},x^{2},x^{3})$ -- acting in a pseudo-Riemannian spacetime,
and we perform the variation of $I$ -- say $\delta_{g}I$ --
generated by the variation $\delta g_{jk}$, $(j,k=0,1,2,3)$, of
the metric tensor $g_{jk}(x)$ (possibly interacting with
$\varphi(x)$),

\begin{equation} \label{eq:one}
\delta_{g}I = \int_{D} (\ldots)^{jk} \, \delta g_{jk} \, \sqrt{-g}
\,\, \textrm{d}^{4}x \quad,
\end{equation}

-- where $D$ is a fixed spacetime domain -- , the expression
$(\ldots)^{jk}$ is a symmetrical tensor, which represents the
energy-momentum-stress tensor of $\varphi(x)$. This statement has
been \emph{verified} for various fields \cite{1}. And its
\emph{general} validity can be intuitively understood bearing in
mind that $I$ is an action integral, with the Lagrange density of
$\varphi(x)$ as integrand.

\par \emph{We shall prove that the above statement holds also if}
$\varphi(x) \equiv g_{jk}(x)$, thus corroborating a famous (and
debated!) thesis by Lorentz \cite{2} and Levi-Civita \cite{3} --
see also Pauli \cite{4} (and the references therein).

\par The essential merit of the following demonstration is its
\textbf{\emph{independence}} \emph{of the Einstein field
equations} (and of the Bianchi relations).

\vskip1.20cm \noindent \textbf{2.} -- Let $\sqrt{-g}\,\, S \,\,
[g_{jk}(x), g_{jk,m}(x),g_{jk,mn}(x), \ldots]$ be a generic scalar
density which is a function of the metric $g_{jk}(x)$ and of a
finite number of its ordinary derivatives \cite{5}. We do
\textbf{\emph{not}} assume that $\sqrt{-g}\,\, S$ is a Lagrange
density, and therefore the integral

\begin{equation} \label{eq:two}
\mathcal{J} = \int_{D} S \, \sqrt{-g} \,\, \textrm{d}^{4}x
\end{equation}

is \textbf{\emph{not}} an action integral. We have:

\begin{equation} \label{eq:three}
\delta_{g}\, \mathcal{J} = \int_{D} \frac{\delta (S \,
\sqrt{-g})}{\delta g_{jk}} \, \delta g_{jk} \, \textrm{d}^{4}x
\quad;
\end{equation}

the variational derivative $\delta (S \, \sqrt{-g}) / \delta
g_{jk}$ is equal to

\begin{equation} \label{eq:four}
\frac{\partial (S \, \sqrt{-g})}{\partial g_{jk}} -
\frac{\partial}{\partial x^{m}} \left[\frac{\partial (S \,
\sqrt{-g})}{\partial g_{jk,m}}\right] +
\frac{\partial^{2}}{\partial x^{m} x^{n}} \left[\frac{\partial (S
\, \sqrt{-g})}{\partial g_{jk,mn}}\right] - \, \ldots \quad ;
\end{equation}

putting $\delta (S \, \sqrt{-g}) / \delta g_{jk} := P^{jk} \,
\sqrt{-g}$, we can write:

\begin{equation} \label{eq:threeprime}
\delta_{g}\, \mathcal{J} = \int_{D} P^{jk} \, \sqrt{-g} \,\,
\delta g_{jk} \, \textrm{d}^{4}x \tag{3$'$} \quad.
\end{equation}

Let us now consider the \emph{particular} $\delta g_{jk}$ -- say
$\delta^{*} g_{jk}$ --, which is generated by an infinitesimal
change of the co-ordinates $x$:

\begin{equation} \label{eq:five}
x\,'^{j} = x^{j} + \varepsilon^{j}(x) \quad ;
\end{equation}

we assume that $\varepsilon^{j}(x)$ is zero on the bounding
surface $\partial D$. The corresponding variation of $\mathcal{J}$
-- say $\delta^{*}_{g}\mathcal{J}$ -- will be equal to
\emph{zero}, because $\mathcal{J}$ is an invariant.

\par We have:

\begin{equation} \label{eq:six}
g_{mn}(x) = \frac{\partial x\,'^{j}}{\partial x^{m}} \, \,
\frac{\partial x\,'^{k}}{\partial x^{n}} \, \, g\,'_{jk} (x\,')
\quad ,
\end{equation}

and we consider the $\delta^{*} g_{jk}$ for \emph{fixed} values of
the coordinates, \emph{i.e.}:

\begin{equation} \label{eq:sixprime}
\delta^{*} g_{jk} :=  g\,'_{jk}(x\,') - g_{jk}(x\,') =
g\,'_{jk}(x\,') - g_{jk}(x) - g_{jk,s}(x) \, \varepsilon^{s}
\tag{6$'$} \quad.
\end{equation}

It follows immediately from eqs.(\ref{eq:five}), (\ref{eq:six}),
(\ref{eq:sixprime}) that

\begin{equation}\label{eq:seven}
\delta^{*} g_{mn} = -g_{mn,j} \, \varepsilon^{j} - g_{mj} \,
\varepsilon^{j}_{,\,n} - g_{nj} \, \varepsilon^{s}_{,\,m} \quad ;
\end{equation}

from eq.(\ref{eq:threeprime}) we get:

\begin{eqnarray}\label{eq:eigth}
\delta^{*}_{g} \mathcal{J} & = & \int_{D} P^{mn} \, \sqrt{-g} \,
\delta^{*}g_{mn}  \, \textrm{d}^{4}x =
\nonumber\\
& & = \int_{D} P^{mn} \left( -g_{m};  \varepsilon^{j}_{,\,n} -
g_{nj} \, \varepsilon^{j}_{,\,m} - g_{mn,j} ;  \varepsilon^{j}
\right) \, \sqrt{-g} \, \textrm{d}^{4}x =
\nonumber\\
& & =\int_{D} \left[ 2 \, (P_{j}^{n} \, \sqrt{-g})_{, \, n} -
g_{mn,j} \, P^{mn} \, \sqrt{-g}) \right] \, \varepsilon^{j} \,
\textrm{d}^{4}x
= \nonumber\\
& & =2 \, \int_{D} P_{j:\, m}^{m} \, \varepsilon^{j} \, \sqrt{-g}
\, \textrm{d}^{4}x = 0 \quad ,
\end{eqnarray}

if the colon denotes a covariant derivative; in the last passage
we use the following property of any symmetrical tensor $S^{mn}$:

\begin{equation} \label{eq:eigthprime}
S_{j:\,m}^{m} \, \sqrt{-g} = \left( S_{j}^{n} \, \sqrt{-g}
\right)_{,n} - \frac{1}{2} \, g_{mn,j} \, S^{mn} \, \sqrt{-g}
\tag{8$'$} \quad.
\end{equation}

Accordingly:

\begin{equation} \label{eq:nine}
P_{j:\,m}^{m} = 0 \quad ; \quad (j=0,1,2,3) \quad .
\end{equation}

 \vskip1.20cm \noindent\textbf{3.} -- The result (\ref{eq:nine})
 has a mere mathematical interest. It becomes physically
 significant when $\mathcal{J}$ is the action integral, say $A$,
 given by

\begin{equation} \label{eq:ten}
A = \int_{D} R \, \sqrt{-g} \, \textrm{d}^{4}x \quad,
\end{equation}

where $R=R^{jk}g_{jk}$ is the Ricci scalar. We shall not use the
fact that the $g_{jk}$'s are (\emph{a priori}) independent
variables, because we do not wish to deduce from the action $A$
the Einstein field equations.

\par Standard procedures (see, \emph{e.g.}, Hilbert's method in
\emph{Appendix}, \mbox{\boldmath $\beta$})) tell us that

\begin{equation} \label{eq:eleven}
\delta_{g}A = \int_{D}\left( R^{jk} - \frac{1}{2} \, g^{jk} \,
R\right) \, \sqrt{-g} \, \, \delta g_{jk} \, \textrm{d}^{4}x
\quad;
\end{equation}

the analogue of eq.(\ref{eq:eigth}) is:

\begin{equation} \label{eq:twelve}
\delta^{*}_{g}A = 2 \, \int_{D}\left( R_{j}^{k} - \frac{1}{2} \,
\delta^{j}_{k} \, R\right)_{:\, k} \, \varepsilon^{j} \, \sqrt{-g}
\, \,
 \textrm{d}^{4}x = 0 \quad ,
\end{equation}

from which:

\begin{equation} \label{eq:thirteen}
\left( R^{jk} - \frac{1}{2} \, g^{jk} \, R\right)_{:\, k} = 0
\quad, \quad (j=0,1,2,3) \quad .
\end{equation}

Thus, quite independently of the field equations, we see that the
\emph{symmetrical} tensor $R^{jk}-(1/2)g^{jk}R$ satisfies four
\emph{conservation equations}. Of course, eqs.(\ref{eq:thirteen})
are identically satisfied by virtue of Bianchi relations, but the
above method -- which is essentially due to the conceptions of
Emmy Noether \cite{6} -- evidences the conservative property of
$R^{jk}-(1/2)g^{jk}R$, and attributes it the nature of an
energy-momentum-stress tensor. Properly speaking,
$[R^{jk}-(1/2)g^{jk}R]/\kappa$, if $\kappa$ is the Newton-Einstein
gravitational constant, represents the Einsteinian energy tensor,
as it was emphasized by Lorentz \cite{2} and Levi-Civita \cite{3}.
And the fact that this tensor is a function \emph{only} of the
potential $g^{jk}$ implies that it is \emph{the unique}
energy-momentum-stress tensor of the gravitational field.

 \vskip1.20cm \noindent\textbf{4.} --  The fact that
 $[R^{jk}-(1/2)g^{jk}R]/\kappa$ is \emph{the true}
 energy-momentum-stress tensor of the gravitational field has a
 very important consequence \cite{3}: the mathematical undulatory
 solutions of the equations $R^{jk}-(1/2)g^{jk}R = 0 = R^{jk}$ are
 quite devoid of physical meaning, because they do not transport
 energy, momentum, stress. This was the \emph{first}
 demonstration  of the physical non-existence of the gravitational
 waves. Quite different demonstrations have been given in recent
 years, see \emph{e.g.} \cite{7}, and references therein.

 \par In his fundamental memoir \cite{3}, Levi-Civita proved also
 the nature of mere \emph{mathematical fiction} (Eddington
 \cite{8}) of the well-known pseudo energy-tensor of the metric
 field $g_{jk}$. --

\vskip0.50cm \noindent A useful discussion with Dr. T. Marsico is
gratefully acknowledged.

\vskip1.60cm
\begin{center}
\noindent \small \emph{\textbf{APPENDIX}}
\end{center} \normalsize

\vskip0.40cm \noindent \mbox{\boldmath $\alpha$}) The full
illogicality of the notion of pseudo energy-tensor can be seen
also in the following way. The usual definition of this pseudo
tensor is:

\begin{equation} \label{eq:A1}
\sqrt{-g}  \,\, t_{m}^{\,\,\,n} \, \stackrel {DEF}{=} \,
\frac{\partial (L \sqrt{-g})}{\partial g_{jk,n}} \, g_{jk,m} -
\delta_{m}^{n} \, L \, \sqrt{-g} \quad ; \tag{A.1}
\end{equation}

the function $L$:

\begin{equation} \label{eq:A2}
L \equiv g^{mn} \left( \Gamma^{s}_{\,mn} \, \Gamma^{r}_{\,sr} -
\Gamma^{r}_{\,ms} \, \Gamma^{s}_{\,nr} \right) \tag{A.2}
\end{equation}

yields the Lagrangean field equations:

\begin{equation} \label{eq:A3}
\frac{\partial (L \sqrt{-g})}{\partial g_{jk}} -
\frac{\partial}{\partial x^{n}} \left[ \frac{\partial (L
\sqrt{-g})}{\partial g_{jk,n}} \right] = 0 \quad . \tag{A.3}
\end{equation}

Now, the left-hand side of (\ref{eq:A3}) is \textbf{\emph{not}}
equal to

\begin{equation} \label{eq:A4}
- \left( R^{jk} - \frac{1}{2} g^{jk} R \right) \sqrt{-g}
 \tag{A.4}
\end{equation}

as it is commonly affirmed. Indeed:

\begin{itemize}
\item[\emph{i})] A non tensor entity cannot be equal to a tensor
density --

\item[\emph{ii})] The above affirmed equality has its origin in a
``negligence'': in the customary variational deduction of the
Einstein field equations the variation of $\int_{D} R \sqrt{-g} \,
\textrm{d}^{4}x $ is ``reduced'' to the variation of $\int_{D} L
\sqrt{-g} \, \textrm{d}^{4}x $. But in his ``reduction'' two
perfect differentials in the integrand have been omitted, because
on the boundary $\partial D$ the variations of the $g_{jk}$ and of
their first derivatives are zero (by assumption): this omission
\emph{destroys} the tensor-density character of the initial
expressions. --
\end{itemize}

\noindent \mbox{\boldmath $\beta$}) It is likely that the pseudo
energy-tensor would not have been invented if the authors had
followed Hilbert's procedure \cite{9}. This Author started from
the fact that (with our previous notations) the explicit
evaluation of the variational derivative $\delta (R \sqrt{-g}) /
\delta g^{mn}$ gives the following Lagrangean expressions:

\begin{equation} \label{eq:A5} \tag{A.5}
\frac{\partial (R \sqrt{-g})}{\partial g^{mn}} -
\frac{\partial}{\partial x^{k}} \left[ \frac{\partial (R
\sqrt{-g})}{\partial g^{mn}_{,\,k}} \right] + \frac{\partial
^{2}}{\partial x^{k} x^{l}} \left[ \frac{\partial (R
\sqrt{-g})}{\partial g^{mn}_{,\,kl}} \right] \quad ;
\end{equation}

Hilbert wrote: ``$\ldots$ specializiere man zun\"{a}chst das
Koordinatensystem so, da\ss {} f\"{u}r den betrachteten Weltpunkt
die $g^{mn}_{,s}$ s\"{a}mtlich verschwinden.''. \emph{I.e.}, he
chose a \emph{local} coordinate-system for which the \emph{first}
derivatives of $g^{mn}$ are equal to zero. Thus, only the first
term of (\ref{eq:A5}) gives a non-zero contribution, and we have
that (\ref{eq:A5}) is equal to

\begin{equation} \label{eq:A6}
\sqrt{-g}  \, \left( R_{mn} -  \frac{1}{2} \, g_{mn} \, R \right)
\quad .\tag{A.6}
\end{equation}

There is no room in this procedure for false (pseudo) tensor
entities.

\end{document}